\documentclass[%
 pra, twocolumn,
 superscriptaddress,
 shortbibliography,
 aps,
 ]{revtex4-2}

\usepackage[pdftex]{hyperref,color,graphicx}
\usepackage{amsfonts,amssymb,amsmath}
\usepackage[separate-uncertainty=false]{siunitx}
\usepackage[english]{babel}
\usepackage[utf8]{inputenc}
\usepackage[T1]{fontenc}

\usepackage{graphicx}
\usepackage[space]{grffile}
\usepackage{latexsym}
\usepackage{textcomp}
\usepackage{longtable}
\usepackage{multirow,booktabs}
\usepackage{url}
\hypersetup{colorlinks=false,pdfborder={0 0 0}}
\usepackage[english]{babel}
\usepackage{dcolumn}
\usepackage{bm}
\usepackage{units}
\usepackage{upgreek}
\usepackage{color}
\usepackage{cancel}
\usepackage{hyperref}
\usepackage{braket}
\usepackage{amsmath}
\newcommand\numberthis{\addtocounter{equation}{1}\tag{\theequation}}

\makeatletter
\renewcommand{\p@subsection}{}
\renewcommand{\p@subsubsection}{}
\makeatother
\usepackage[normalem]{ulem}

\begin{document}

\title{Injection locking of a levitated optomechanical oscillator for precision force sensing}

\author{Siamak Dadras}
\affiliation{The Institute of Optics, University of Rochester, Rochester, NY 14627, USA}
\affiliation{Center for Coherence and Quantum Optics, University of Rochester, Rochester, NY 14627, USA}

\author{Robert M. Pettit}
\affiliation{Institute for Research in Electronics and Applied Physics, and Joint Quantum Institute, University of Maryland, College Park, MD 20742, USA}

\author{Danika R. Luntz-Martin}
\affiliation{Center for Coherence and Quantum Optics, University of Rochester, Rochester, NY 14627, USA}
\affiliation{Department of Physics and Astronomy, University of Rochester, Rochester, NY 14627, USA}

\author{Kewen Xiao}
\affiliation{School of Physics and Astronomy, Rochester Institute of Technology, Rochester, NY 14623, USA}  

\author{M. Bhattacharya}
\affiliation{Center for Coherence and Quantum Optics, University of Rochester, Rochester, NY 14627, USA}
\affiliation{School of Physics and Astronomy, Rochester Institute of Technology, Rochester, NY 14623, USA}

\author{A. Nick Vamivakas}
\email{nick.vamivakas@rochester.edu}
\affiliation{The Institute of Optics, University of Rochester, Rochester, NY 14627, USA}
\affiliation{Center for Coherence and Quantum Optics, University of Rochester, Rochester, NY 14627, USA}
\affiliation{Department of Physics and Astronomy, University of Rochester, Rochester, NY 14627, USA}

\begin{abstract}
We report on the injection locking of an optically levitated nanomechanical oscillator (a silica nanosphere) to resonant intensity modulations of an external optical signal. We explore the characteristic features of injection locking in this system, e.g. the phase pull-in effect and the injection-induced reduction of the oscillation linewidth. Our measurements are in good agreement with theoretical predictions and deepen the analogy of injection locking in levitated optomechanical systems to that in optical systems (lasers). By measuring the force noise of our feedback cooled free-running oscillator, we attain a force sensitivity of $\sim23~\rm{zN}/\sqrt{\rm{Hz}}$. This can readily allow, in fairly short integration times, for tests of violations of Newtonian gravity and searching for new small-scale forces. As a proof of concept, we show that the injection locking can be exploited to measure the forces optically induced on levitated nanoparticles, with potential applications in explorations of optical binding and entanglement between optically coupled nanomechanical oscillators.    
\end{abstract}

\maketitle

\section{INTRODUCTION}\label{sec:intro}

A great incentive in the development of nano- and microscale optomechanical oscillators has been their extreme sensitivity in detecting infinitesimal external force \cite{tao2014single, ranjit2016zeptonewton, hempston2017force, hebestreit2018sensing, moser2013ultrasensitive, gavartin2012hybrid, reinhardt2016ultralow, li2007ultra, doolin2014multidimensional, miao2012microelectromechanically}, displacement \cite{doolin2014multidimensional, miao2012microelectromechanically, teufel2009nanomechanical, wilson2015measurement}, torque \cite{kim2016approaching, hoang2016torsional, ahn2020ultrasensitive, kuhn2017optically}, acceleration \cite{krause2012high, guzman2014high, wisniewski2020optomechanical}, charge \cite{cleland1998nanometre, moore2014search}, and added mass \cite{chaste2012nanomechanical, liu2013sub}. Notably, sub-attonewton force sensing paves the way for a variety of applications such as magnetic resonance force microscopy and imaging \cite{rugar2004single, degen2009nanoscale}, detecting gravitational waves \cite{arvanitaki2013detecting}, and searching in short ranges for non-Newtonian gravity \cite{geraci2008improved, geraci2010short}, surface forces \cite{geraci2015sensing, winstone2018direct, canaguier2011casimir, nie2012effect}, and interactions associated with dark energy \cite{rider2016search}. While there has been impressive progress in the development of resonant solid-state force sensors such as dielectric microcantilevers \cite{tao2014single} and carbon nanotubes \cite{moser2013ultrasensitive}, these devices typically operate at cryogenic temperatures to improve their thermal-noise-limited force sensitivity. Cryogenic cantilever and nanotube oscillators have reached sensitivities of  $\sim1~\rm{aN}/\sqrt{\rm{Hz}}$ \cite{tao2014single} and $\sim10~\rm{zN}/\sqrt{\rm{Hz}}$ \cite{moser2013ultrasensitive} respectively, and room-temperature solid-state sensors have been realized with sensitivities in the range of $10-500~\rm{aN}/\sqrt{\rm{Hz}}$ \cite{gavartin2012hybrid, miao2012microelectromechanically, doolin2014multidimensional, reinhardt2016ultralow, li2007ultra}.

In contrast to these mechanically clamped oscillators, mesoscopic particles levitated in high vacuum (HV) are recognized as low-dissipation optomechanical oscillators due to their minimal thermal contact to the environment \cite{li2011millikelvin, gieseler2012subkelvin, chang2010cavity, millen2015cavity}. The levitated oscillator's low mass and excellent environmental isolation in HV has allowed such systems to achieve, at room temperature, similar or better quality factors and force sensitivities than their tethered counterparts \cite{ranjit2016zeptonewton, hempston2017force, hebestreit2018sensing, gieseler2013thermal, millen2020optomechanics}. With these prospects, levitated optomechanics has become the backbone of many state-of-the-art experiments, ranging from sensing and metrology \cite{millen2020optomechanics} to the study of phonon lasers \cite{pettit2019optical, vahala2009phonon} and hybrid systems with mechanical and spin degrees of freedom \cite{neukirch2015multi, pettit2017coherent}. It may also provide a remarkable platform for exploring quantum mechanics at the macroscale, such as study of macroscopically separated superposition states \cite{arndt2014testing, romero2011quantum}, tests of collapse models \cite{romero2011quantum, bassi2013models}, matter-wave interferometry \cite{hornberger2012colloquium, bateman2014near}, and the Schr\"{o}dinger-Newton equation \cite{grossardt2016optomechanical}. 

Precision force sensing with optically trapped dielectric oscillators has been implemented or proposed for a variety of schemes. Ranjit {\it et al.} \cite{ranjit2016zeptonewton} and Hempston {\it et al.} \cite{hempston2017force} showed the capacity of charged, feedback-cooled micro- and nanospheres to detect Coulomb forces from oscillating electric fields, and achieved force sensitivities of $1.6~\rm{aN}/\sqrt{\rm{Hz}}$ and $32~\rm{zN}/\sqrt{\rm{Hz}}$ respectively. Gieseler {\it et al.} used a feedback-cooled silica nanoparticle with $20~\rm{zN}/\sqrt{\rm{Hz}}$ sensitivity to detect a periodic optical force gradient induced by a low frequency modulation of the trapping potential \cite{gieseler2013thermal}. In the context of detecting surface forces, Rider {\it et al.} reported a sensitivity of $20~\rm{aN}/\sqrt{\rm{Hz}}$ for a silica microsphere trapped in close proximity to an oscillating Au-coated silicon cantilever \cite{rider2016search}. Diehl {\it et al.} trapped a silica nanoparticle at a subwavelength distance from a SiN membrane with envisioned implications in the study of short-range interactions \cite{diehl2018optical}. Winstone {\it et al.} observed distortion of the trapping potential as a surface-induced effect on a charged silica particle and reported a sensitivity of $\sim80~\rm{aN}/\sqrt{\rm{Hz}}$ for this system \cite{winstone2018direct}. Magrini {\it et al.} estimated $\sim10~\rm{zN}/\sqrt{\rm{Hz}}$ sensitivity in the near-field coupling of a levitated nanoparticle to a photonic crystal cavity \cite{magrini2018near}. In the same framework, Geraci {\it et al.} proposed sensing short-range forces using a matter-wave interferometer, in which a falling nanosphere released from an optical trap interacts with a mass wall in its close proximity \cite{geraci2015sensing}. 

Force sensing with levitated particles extends further to cavity optomechanical systems. Geraci {\it et al.} proposed trapping and cooling a microsphere in the antinode of an optical cavity field and reasoned that such a high-$Q_m$ system may lead to yN force sensitivity with potential applications in the detection of short-range forces \cite{geraci2010short}. A later proposal from the same group described a cavity-based tunable resonant sensor to detect gravitational waves using optically trapped and cooled microspheres or microdisks \cite{arvanitaki2013detecting}. The detection of static forces was considered by Hebestreit {\it et al.} through the measurement of force-induced displacements on a particle in free--fall after being released from an optical trap, which enabled gravitational and electrostatic force resolution at the 10 aN level \cite{hebestreit2018sensing}. Despite the unprecedented sensitivities of levitated force sensors, their accuracy is usually undermined by uncertainties in the effective mass of the oscillator. By applying a periodic Coulomb force on a charged nanosphere, Ricci {\it et al.} presented a novel protocol to measure the particle's mass through its electrically driven dynamics \cite{ricci2019accurate}. This improved both precision and accuracy by more than one order of magnitude, potentially enabling paramount advances in the applications of levitated systems as force sensors. In addition to all these, there have been a number of schemes relying on the coupling of external forces to, and measurement via, the orientational degrees of freedom of levitated anisotropic particles. Hoang {\it et al.} \cite{hoang2016torsional} and Ahn {\it et al.} \cite{ahn2020ultrasensitive} reported $\sim10^{-29}$ and $\sim10^{-27}\rm{N\cdot m}/\sqrt{\rm{Hz}}$ torque sensitivities for levitated nonspherical nanoparticles and nanodumbbells with torsional vibration frequencies of $\sim$1 MHz and $\sim$5 GHz respectively. This would, for example, allow for the detection of rotational vacuum friction \cite{zhao2012rotational} and Casimir torque \cite{manjavacas2017lateral} near a surface.

In spite of these diverse realizations of optomechanical force sensors, notably those with levitated dielectric particles, no implementation has been reported on the measurement of forces exerted via injection locking of an external harmonic signal to a levitated nanoparticle. Injection locking, first noted by Huygens in pendulum clocks in 1665 \cite{bennett2002huygens}, is a well known effect in both solid state \cite{kurokawa1973injection} and optical oscillators (lasers) \cite{stover1966locking}. When a free-running, self-sustained oscillator is exposed to a weaker harmonic signal, its phase and frequency can be locked to that of the injected signal if the frequency difference between the two is sufficiently small. This effect has also been observed in an array of levitated systems, including rf-driven Paul-trapped ions \cite{knunz2010injection} and graphene nanoplatelets \cite{coppock2016phase}, as well as optically trapped and driven nanospheres \cite{gieseler2014nonlinear} and silicon nanorods \cite{kuhn2017optically}. The first enabled the detection of Coulomb forces as low as $\sim$5 yN (largely due to the low mass of the ion, though naturally sensitive to electric and magnetic noises) and the last was predicted to detect torques with $\sim$0.25 zN$\cdot$m sensitivity. However, the majority of injection-locked optomechanical systems exploit tethered microcavities such as microtoroids \cite{alaie2016enhancing, bekker2017injection, huang2018injection}, microdisks \cite{zhang2012synchronization, shah2015master, zhang2015synchronization}, photonic cavities \cite{bagheri2013photonic, li2012multichannel}, and integrated chipsets \cite{luan2014integrated}. Injection locking has been performed in these systems for the synchronization of oscillator networks \cite{zhang2012synchronization, shah2015master, zhang2015synchronization, bagheri2013photonic} and controlling the phase and frequency of individual oscillators \cite{alaie2016enhancing, bekker2017injection, huang2018injection, li2012multichannel, luan2014integrated}. Thus injection locking of levitated dielectric particles with an approach to the measurement of optically induced forces is yet to be explored.

Here we demonstrate injection locking of the mechanical oscillations of a trapped and motionally cooled silica (SiO$_2$) nanosphere to the intensity modulations of an external laser. In addition to exploring the characteristic signatures of injection locking in this scheme, we test, as a proof of concept, its force sensing ability with a $\sim100$ zN-scale injected signal. However, the $\sim23~\rm{zN}/\sqrt{\rm{Hz}}$ sensitivity of our feedback-cooled nanoparticle, suggests its capacity of detecting $\sim1$ zN-scale forces in fairly short integration times. We measure the injected force from the oscillation amplitude by calibrating the system to the amplitude associated with the force noise in the absence of an injected signal \cite{ranjit2016zeptonewton}. Our scheme is highly versatile due to its room-temperature control over the thermal noise in HV and its cavity-free nature of the dipole trap. The latter enables a wide-range frequency tunability of our oscillator via tuning the trapping laser intensity \cite{pettit2019optical}, facilitating its locking to an injected signal of a desired frequency. The versatility of our approach is also due the fact that, unlike in other Coulomb-force-based levitated systems, we measure optically induced forces on charge-free particles, precluding effects of electronic and magnetic noises on the particle's dynamics. All these enable ultrasensitive force measurements in a wide frequency range without resorting to cryogenic environment and/or additional arrangements to accommodate rf, acoustic or electro-optic perturbations. Our approach to the measurement of forces induced by oscillating optical potentials can be, for example, advantageous for explorations of optical binding \cite{arita2018optical} and entanglement \cite{rudolph2020entangling} between two oscillating particles, mediated by the scattered light from one particle coupled to another.  
\section{RESULTS}
\label{sec:results}
\subsection{Injection locking}
\begin{figure}
\centering
\includegraphics[scale=0.32]{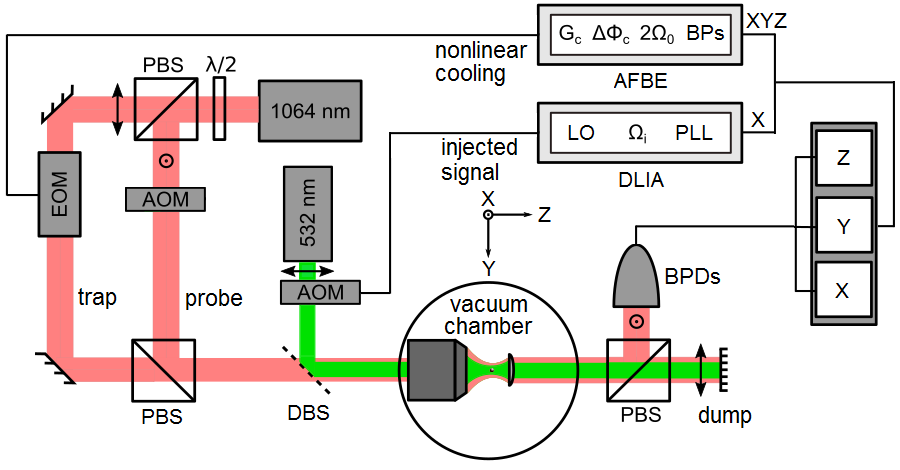}
\caption{Experimental setup for injection locking. AOM -- acousto-optic modulator, EOM -- electro-optic modulator, AFBE -- analog feedback electronics, BPs -- three bandpass filters, $2\Omega_0$ -- three frequency doublers, $\Delta\Phi_{\rm{c}}$ -- three phase shifters, $\rm{G_c}$ gain-tunable summing amplifier, DLIA -- digital lock-in amplifier, PLL -- phase-locked loop, LO -- local oscillator, BPDs -- balanced photodiodes and their associated optics for split-detecting the probing beam in three orthogonal directions, DBS -- dichroic beam splitter, PBS -- polarizing beam splitter. The trapping beam is feedback modulated via the EOM, while the probe beam is separately controlled by an AOM. Forward-scattered probe beam is collected in a network of BPDs from which the 3--axis parametric feedback cooling and the 1--axis injection locking signals are separately derived. The injection laser is intensity modulated with an additional AOM driven by a LO that generates the harmonic modulating signal.}
\label{fig:setup}
\end{figure}

The experimental apparatus is constructed based on a free-space optical trapping and feedback center-of-mass (c.o.m) cooling of a fused silica nanosphere in HV as schematically shown in Fig.~\ref{fig:setup}. Trapping is achieved by tightly focusing a 1064 nm linearly polarized laser beam on the particle. To cool the particle, the probe light scattered from the trapped particle is split-detected in orthogonal directions and processed in a series of analog feedback electronics to derive feedback signals that nonlinearly slow down the particle's c.o.m motion in 3D (see Supplementary Information for details). This feedback enables us to maintain the particle in the trap under HV, where the damping due to residual gas molecules is significantly reduced. Injection locking is realized by introducing a weak 532 nm laser onto our levitated nanosphere. This additional laser exerts a force on the particle via intensity gradient and optical scattering \cite{neukirch2015nano}, and modulating its intensity at a frequency close to that of the particle's oscillation (in $x$ axis here, see Fig.~\ref{fig:setup} for the system's coordinates) produces an injection signal the particle's oscillation phase can lock to. This modulation is achieved separately using a phase-locked loop and a local oscillator in a digital lock-in amplifier, and is distinct from the feedback process. This local oscillator is then used as a phase reference for the measurement of particle's dynamics with respect to the modulation.

With the detuning $\delta=\Omega_i-\Omega_0$ between the frequencies of the injected signal and the free-running oscillator, one can define a characteristic locking range $-\Omega_m\leq\delta\leq+\Omega_m$, over which the external modulation can be detuned and still cause the oscillator to maintain a fixed phase difference of 

\begin{equation}
\Delta\phi=\sin^{-1}(-{\delta}/{\Omega_m})
\label{eqn:adler_fixed}
\end{equation}
with the injected signal (see Supplementary Information for details). In contrast, there is no fixed phase relationship outside of the locking range ($|\delta|>\Omega_m$), where

\begin{equation}
\Delta\phi(t) = 2\tan^{-1}\left[-\frac{\Omega_{b}}{\delta}\tan\left(\frac{\Omega_{b}(t-t_{0})}{2}\right)-\frac{\Omega_m}{\delta}\right]
\label{eqn:adler_time}
\end{equation}
oscillates between $\pm\pi$ and never reaches a steady state (here $\Omega_{b}=\sqrt{\delta^{2}-\Omega_m^{2}}$ and $t_{0}$ is a constant determined by the initial conditions). As illustrated in Fig.~S1, the period of this oscillation increases as $|\delta|\rightarrow\Omega_m$ and becomes increasingly asymmetric. This asymmetry leads to a characteristic pulling effect on the time-averaged phase of the free-running oscillator. 

\begin{figure}
\centering
\includegraphics[scale=0.2]{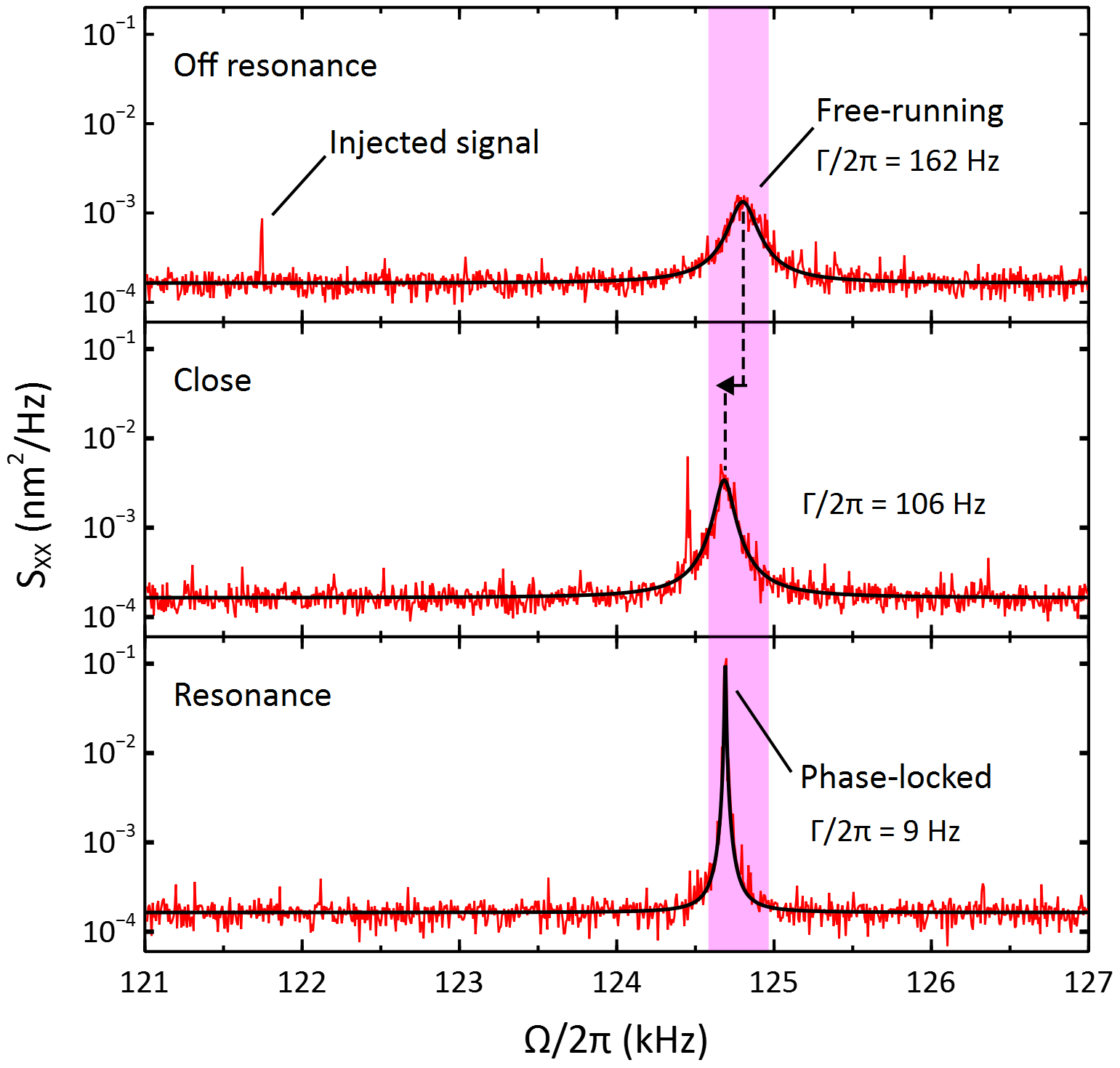}
\caption{Characteristic phase pull-in effect in the injection locking of a levitated nanoparticle. Red shift in the particle's oscillation frequency can be seen in the shaded area when the detuning of the injected signal is swept from far below resonance to inside the locking range. Enhancement of the oscillation amplitude along with a significant reduction in its linewidth can be clearly seen for the locked oscillation. Red: experimental data, black: Lorentzian fit to  data.}
\label{fig:phase_pull}
\end{figure}

Experimental validation of the injection locking and phase-pull effect is illustrated in Fig. \ref{fig:phase_pull} for the $x$ c.o.m degree of freedom of the particle. This figure shows the displacement spectral densities recorded for several detunings from the particle's oscillation. When the injected signal is far outside the locking range (top panel), the free-running oscillation remains intact and a small oscillation component appears at the frequency of the injected signal. As the detuning approaches the locking range (middle panel), the free-running component is pulled towards the injected signal with an apparent decrease in its linewidth and an increase in its amplitude. Inside the locking range (bottom panel), the particle oscillates at a significantly narrower linewidth and a higher amplitude, manifesting signatures of its phase-locking to the injected driving force. 

\begin{figure}
\centering
\includegraphics[scale=0.2]{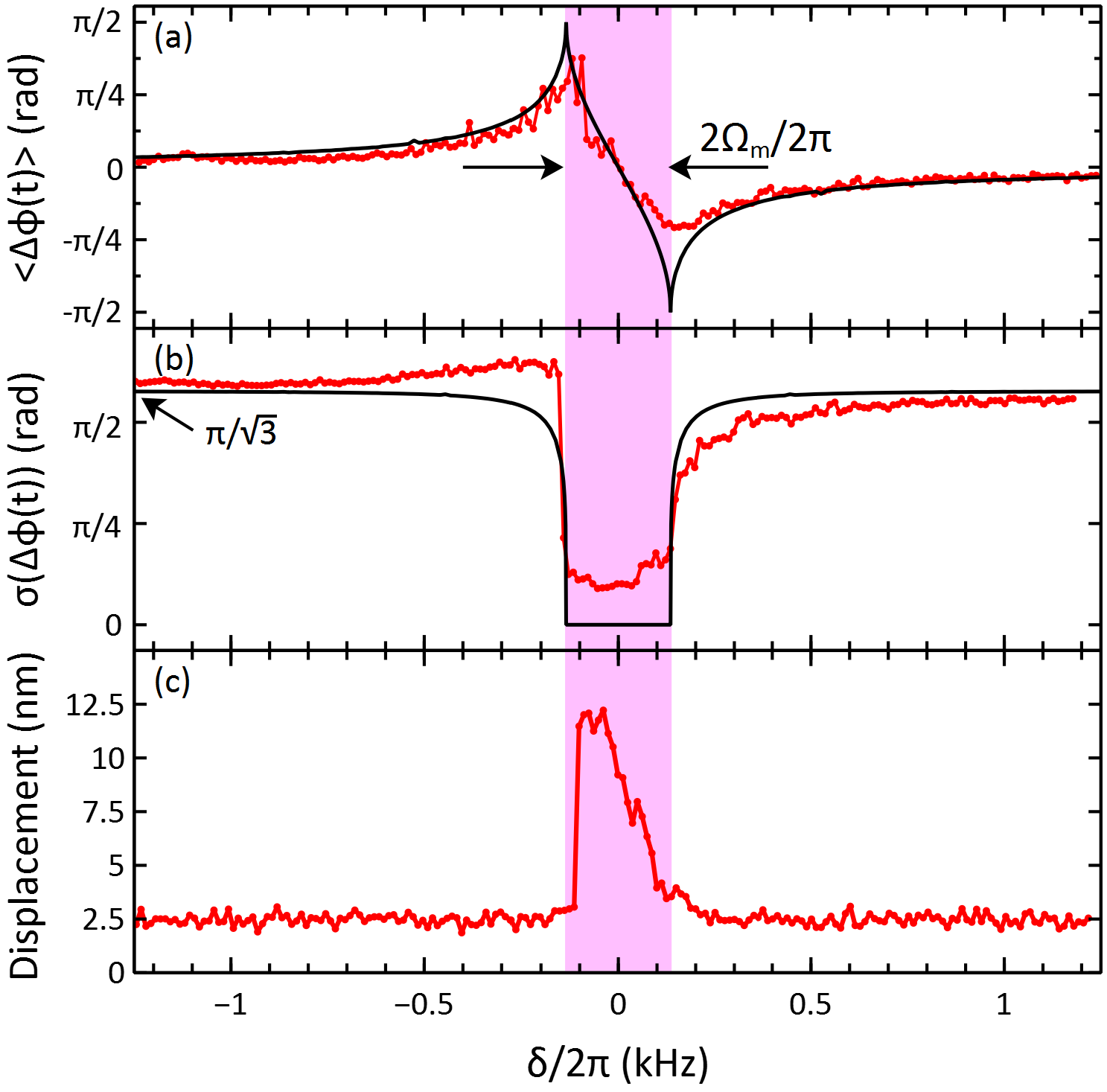}
\caption{(a) Time averaged phase difference between the particle's oscillation and the injected signal when the detuning $\delta$ is scanned over the locking range. Black line shows theoretical fits using Eq.~(\ref{eqn:adler_fixed}) and time averaged Eq.~(\ref{eqn:adler_time}) with $2\Omega_{m}/2\pi=270$ Hz. (b) Standard deviation of the phase difference $\sigma(\Delta\phi(t))$ measured for the same frequency range. The value $\pi/\sqrt{3}$ rad represents the standard deviation under the limit where the phase difference distributes uniformly over the range $\left[-\pi,\pi\right]$. Rapid drop of this quantity to near zero inside the locking range is an evidence of a fairly fixed phase relationship between the particle and the external modulation. Black line shows theoretical fits using standard deviations of Eq.~(\ref{eqn:adler_fixed}) and Eq.~(\ref{eqn:adler_time}) with $2\Omega_{m}/2\pi=270$ Hz. (c) Increase in the RMS displacement when the particle is driven by the injected optical force. The locking range (with the width of $2\Omega_{m}/2\pi$) is shown by the shaded region.}
\label{fig:lockrange_data}
\end{figure}

It is also possible to measure the time-averaged phase difference $\langle\Delta\phi(t)\rangle$. Figure~\ref{fig:lockrange_data}(a) presents this measurement when the frequency of the external signal is swept over the resonance. The phase-pull effect around the locking range is clearly visible in the experimental data and is well matched with Eq.~(\ref{eqn:adler_fixed}) (inside) and time-averaged Eq.~(\ref{eqn:adler_time}) (outside) a locking range of $2\Omega_m/2\pi = 270$ Hz. The phase locking behavior is further evidenced by the variation of the phase difference standard deviation, $\sigma(\Delta\phi(t))$, over the same frequency range (see Fig.~\ref{fig:lockrange_data}(b)). As $\delta$ enters the locking range, the measured $\sigma(\Delta\phi(t))$ drops from $\pi/\sqrt{3}$ rad to $0.09\pi$ rad. For $|\delta| \gg \Omega_m$, a standard deviation of $\pi/\sqrt{3}$ rad is expected for $\Delta\phi(t)$, as this parameter will distribute uniformly over the range $[-\pi,\pi]$. For an ideal locked oscillator ($|\delta|<\Omega_m$), where there is no time dependence in the phase difference $\Delta\phi$, the variance of $\Delta\phi$ would be zero. The deviation of the measured quantities in Figs.~\ref{fig:lockrange_data}(a) and \ref{fig:lockrange_data}(b) from their theoretical predictions is likely due to a combination of factors, namely the thermal fluctuations in the system, imprecision in tracking the oscillator's phase and frequency in the feedback loops, and the home-built nonlinear cooling electronics. The effect of the phase-locking is also clear on the RMS displacement of the particle (Fig.~\ref{fig:lockrange_data}(c)) as it oscillates at a higher (about five times greater) amplitude due to the optical pressure associated with the injected signal.   

\subsection{Force sensing}
Injection locking can be used to measure the forces induced by small optical potentials oscillating at frequencies that the particle's oscillation can lock to. As can be inferred from Fig.~\ref{fig:lockrange_data}(c), one can perform this measurement using a correlation between the amplitude of the phase locked oscillation and the optical force giving rise to this amplitude. We determine this correlation by calibrating the oscillation amplitude with the force noise in the absence of an injected signal \cite{ranjit2016zeptonewton}. The force noise, i.e. the minimum detectable force for such a system is $F_{\rm{min}}=S_F^{1/2}(\Omega_0)b^{1/2}$, with $S_F(\Omega)$ and $b$ being the force noise spectral density and the measurement bandwidth. The force spectral density itself is correlated with the measured displacement spectral density by $S_{xx}(\Omega)=|\chi_m(\Omega)|^2 S_F(\Omega)+S_{xx}^{\rm{imp}}$. Here, $\chi_m(\Omega)=1/\left[m(\Omega_0^2-\Omega^2-i\Gamma\Omega)\right]$ is the optomechanical susceptibility with $m$ the particle's mass, and $\Gamma=\Gamma_0+\delta\Gamma$ the measured damping rate, which is the sum of residual gas (see Eq. S4) and feedback damping rates respectively. Experimentally, the Lorentzian profile of the displacement noise appears on top of a background imprecision noise floor $S_{xx}^{\rm{imp}}$ which accounts for the photon collection efficiency, the feedback electronic noise, the splitting into separate detection paths, optical losses, and the detectors' quantum efficiency. We subtract this background from the total, thus the oscillation amplitude in the subsequent discussion refers to the Lorentzian profile only. For a trapped and feedback cooled particle, the force spectral density at $\Omega=\Omega_0$ will be \cite{rodenburg2016quantum, jain2016direct}
\begin{align*}
S_F(\Omega_0) &= S_t + S_s + S_f(\Omega_0)\\
&=2m\Gamma_0 k_B T_{\rm{eff}} + \frac{2\hbar P_{s}}{5c\lambda} + \frac{3m\hbar \Omega_0 \delta\Gamma^2}{4\chi^2 \Phi}, \numberthis
\label{eqn:sensitivity}
\end{align*}  
where the $S$ terms with their respective expressions denote the contributing sources of noise, i.e. thermal noise $S_t$, shot noise $S_s$, and feedback backaction $S_f$. Here, $k_B$ is Boltzmann's constant, $T_{\rm{eff}}$ is the c.o.m temperature under feedback cooling, and $\lambda$ and $P_{s}$ are the wavelength and scattered power (see Eq. S5) of the trap laser. In addition, $\chi$ ($=10^{-7}$) and $\Phi$ ($=5.2\times10^{14}$ photon/sec) are respectively the scaled optomechanical coupling and the average detected flux of probe photons \cite{rodenburg2016quantum}.

We perform the noise calculation with a nominal set of parameters for a feedback cooled oscillator. This includes $R=73.2$ nm radius and $m=3.6$ fg mass of the particle, $P=2.2\times 10^{-6}$mbar pressure of the chamber which gives a gas damping rate of $\Gamma_0/2\pi=0.0105$ Hz. The trap laser power is $P_0=80$ mW, which with a focusing objective of $\rm{NA}=0.8$ gives a scattered power of $P_{s}=32~\mu$W and an oscillation frequency of $\Omega_0/2\pi=128$ kHz. This results in a mechanical quality factor of $Q_m=\Omega_0/\Gamma_0=1.22\times10^{7}$. In addition, the feedback damping rate measured from the particle's displacement spectral density is $\delta\Gamma/2\pi=550$ Hz, which yields a c.o.m temperature of $T_{\rm{eff}}=23.9$ mK for this particle. With these parameters, we estimate the corresponding noise contributions to be $S_t^{1/2}=0.16$, $S_s^{1/2}=2.06~\rm{zN/\sqrt{Hz}}$, and $S_f^{1/2}=23.03$  giving a total force sensitivity of $S_F^{1/2}=23.12~\rm{zN/\sqrt{Hz}}$.    
\begin{figure}
	\centering
	\includegraphics[scale=0.14]{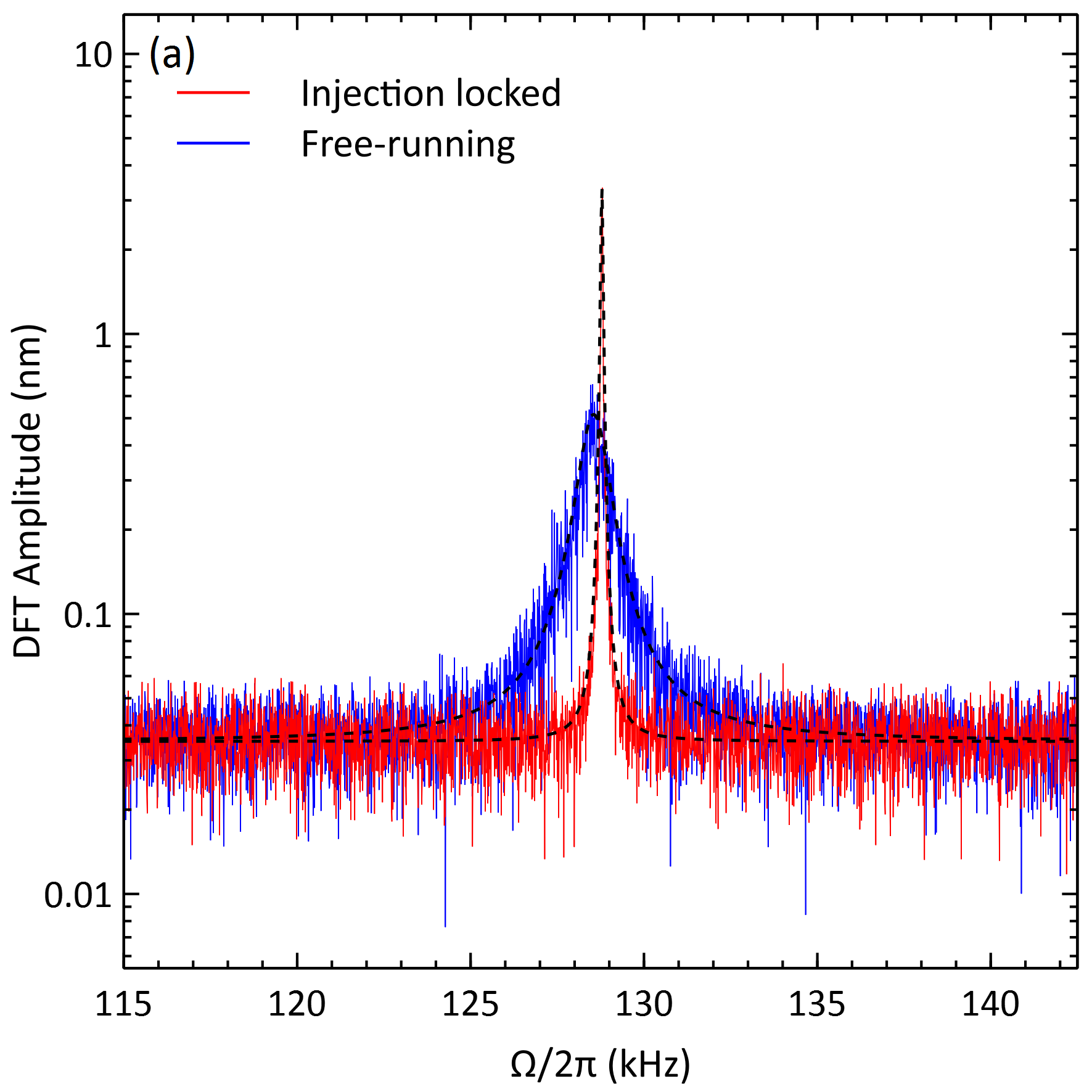}
	\hspace{0.4 cm} \includegraphics[scale=0.14]{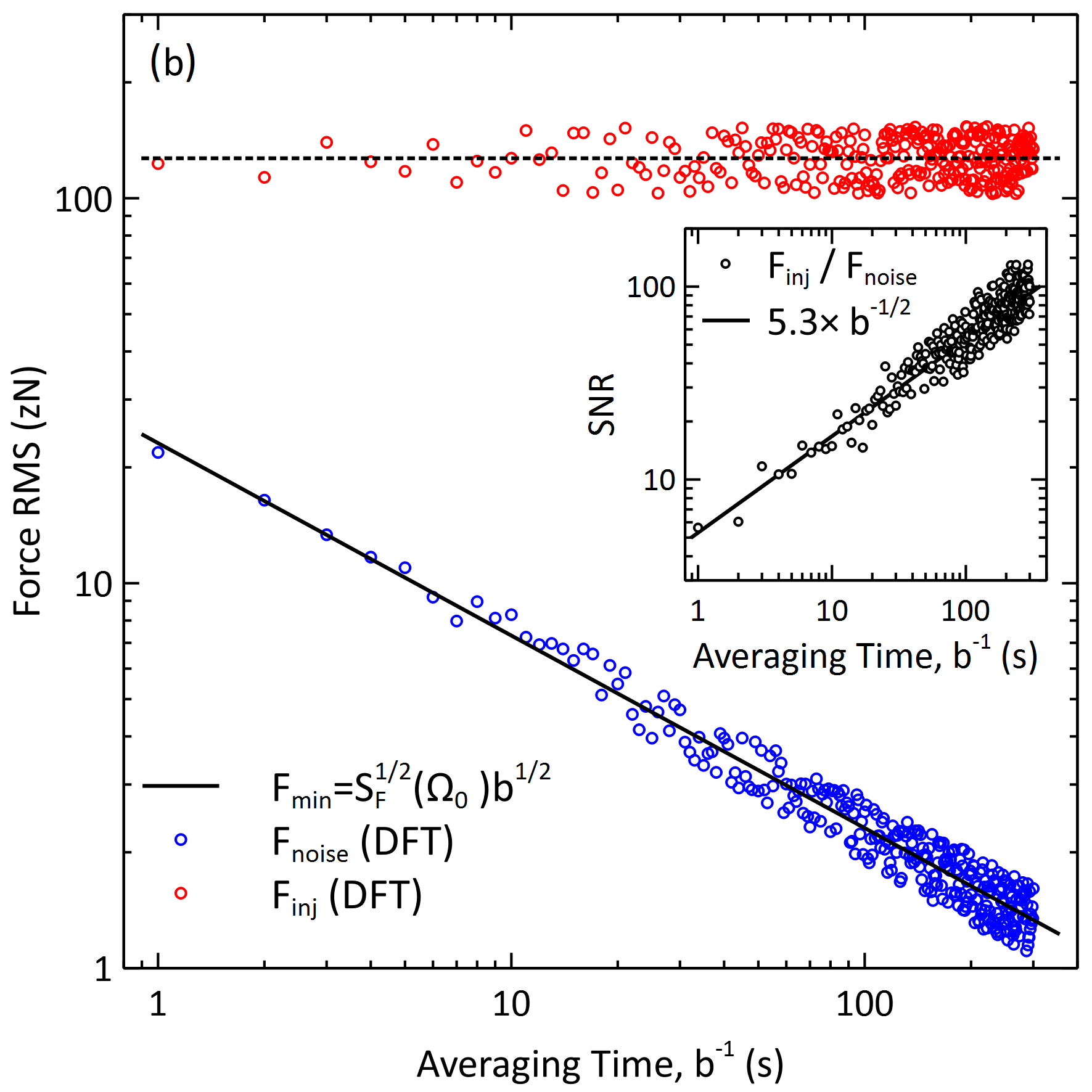}
	\caption{(a) DFT signals of typical free-running (blue) and phase-locked (red) oscillations after $t=1$ sec integration times. An apparent amplified oscillation at a significantly narrower linewidth can be seen for the case of the injection locked oscillation. (b) Forces on the 73.2 nm radius feedback-cooled ($T_{\rm{eff}}=23.9$ mK) nanoparticle in the absence (blue) and presence (red) of a 127 zN injected signal. As expected, the force noise magnitude averages down with the measurement time as $\propto t^{-1/2}$. Force calibration via the noise DFT measurement is used to determine the magnitude of the injected signal. Inset shows the $\propto t^{1/2}$ improvement in the signal-to-noise ratio, i.e. the ratio of the measured force $F_{\rm{inj}}$ to the force noise $F_{\rm{noise}}$, by increasing the measurement time.}
	\label{fig:force_calibrations}
\end{figure}
In the absence of an injected signal, the estimated force noise of the oscillation will average down by increasing the integration time as $F_{\rm{min}}(t)=S_F^{1/2}t^{-1/2}$. This force is linearly proportional to the discrete Fourier transformation (DFT) amplitude of the oscillation $S_x(\Omega_0,t)$ ($\propto S_F^{1/2}$) averaged for the corresponding integration time. Thus a conversion factor, defined as $C=F_{\rm{min}}(t)/S_x(\Omega_0,t)$, can be used straightforwardly to determine the force of the injected signal by measuring the added amplitude of the average DFT, when the locking is engaged. Figure~\ref{fig:force_calibrations}(a) shows DFT signals of typical free-running (blue) and phase-locked (red) oscillations after $t=1$ sec integration times. An apparent amplified oscillation at a significantly narrower linewidth can be seen for the case of the injection locked oscillation. Variations of the calibrated forces associated with these oscillations are shown in Fig.~\ref{fig:force_calibrations}(b) for up to 300 sec integration times. For the typical signal used here to demonstrate the injection locking-based force sensing, we measure an optical driving force of $127$ zN. However, the steady average-down in the force noise of our free-running oscillator suggests the potential of this system to detect forces as small as $\sim1$ zN in a moderate measurement bandwidth of $b\approx (300~\rm{s})^{-1}$.

\section{Conclusions}	
We have demonstrated injection locking of a levitated nanomechanical oscillator to resonant intensity modulations of an external optical signal. We study the characteristic features of injection locking for this system, e.g the phase pull-in effect and significant reduction in the oscillation linewidth imposed by the injected signal. Our measurements are in good agreement with theoretical predictions for an injection locked system and deepen the analogy of our injection locking of a levitated nanomechanical oscillator to that of optical systems, such as lasers. In addition, by measuring the force noise of our feedback cooled free-running oscillator, we show that our system allows for $\sim$1 zN force sensing in fairly short integration times. The zN-scale sensing ability of our system should readily allow for tests of violations of Newtonian gravity ($\sim$1 aN) and searching for small-scale ($\sim$1 zN) forces in short ranges. Furthermore, as a proof of concept, we report on the adoption of the injection locking in levitated optomechanics in the measurement of the forces induced by oscillating optical potentials. This can pave the way for the explorations of small-scale optically induced forces in, for example, optically bound \cite{arita2018optical} and entangled \cite{rudolph2020entangling} levitated nanoparticles. 

\section*{Acknowledgments}
The authors acknowledge Office of Naval Research awards N00014-17-1-2285 and N00014-18-1-2370.
\bibliography{shortref.bib}

\end{document}